\documentclass[conference]{IEEEtran}
\IEEEoverridecommandlockouts

\usepackage{cite}
\usepackage{amsmath,amssymb,amsfonts}
\usepackage{algorithmic}
\usepackage{graphicx}
\usepackage{textcomp}
\usepackage{xcolor}
\usepackage{xspace}
\usepackage{amsmath}
\usepackage{multirow}
\usepackage{ae,aecompl}
\usepackage[T1]{fontenc}
\usepackage{times}
\usepackage{xcolor, soul}
\sethlcolor{red}
\usepackage{pslatex}
\usepackage{mathrsfs}
\usepackage{float}


\def\BibTeX{{\rm B\kern-.05em{\sc i\kern-.025em b}\kern-.08em
    T\kern-.1667em\lower.7ex\hbox{E}\kern-.125emX}}
\begin{document}
\title{Capturing Chronology and Extreme Values of Representative Days for Planning of Transmission Lines and Long-Term Energy Storage Systems\\

\thanks{This work is part of the project Heuristic Efficient Proxy-based Planning of Integrated Energy Systems (HEPPIE), funded by Réseau de Transport d’Électricité (RTE).}
}
\author{\IEEEauthorblockN{Mojtaba Moradi-Sepahvand}
\IEEEauthorblockA{\textit{Department of Electrical Sustainable Energy} \\
\textit{Delft University of Technology}\\
Delft, The Netherlands\\
m.moradisepahvand@tudelft.nl}
\and
\IEEEauthorblockN{Simon H. Tindemans}
\IEEEauthorblockA{\textit{Department of Electrical Sustainable Energy} \\
\textit{Delft University of Technology}\\
Delft, The Netherlands\\
s.h.tindemans@tudelft.nl}
}
\maketitle
\begin{abstract}
The growing penetration of renewable energy sources (RESs) is inevitable to reach net zero emissions. In this regard, optimal planning and operation of power systems are becoming more critical due to the need for modeling the short-term variability of RES output power and load demand. Considering hourly time steps of one or more years to model the operational details in a long-term expansion planning scheme can lead to a practically unsolvable model. Therefore, a clustering-based hybrid time series aggregation algorithm is proposed in this paper to capture both extreme values and temporal dynamics of input data by some extracted representatives. The proposed method is examined in a complex co-planning model for transmission lines, wind power plants (WPPs), short-term battery and long-term pumped hydroelectric energy storage systems. The effectiveness of proposed mixed-integer linear programming (MILP) model is evaluated using a modified 6-bus Garver test system. The simulation results confirm the proposed model efficacy, especially in modeling long-term energy storage systems.\end{abstract}
\vspace{0.3\baselineskip}
\section{Introduction}

\IEEEPARstart{T}{he} increasing penetration of intermittent renewable energy sources (RESs) makes it necessary to deal with the variability of their output power. Energy storage systems (ESSs) are able to store electrical energy during off-peak time periods and supply load demand in peak or emergency periods. In this regard, ESS can be used as an effective solution to increase the operational flexibility of power systems by minimizing load and RES curtailment and providing both intra- and inter-period arbitrage. ESSs can relieve congestion in transmission lines and defer new investment in both transmission lines and generation units. Therefore, in many recent power system expansion studies, ESSs are incorporated as a solution option to optimize the expansion schemes \cite{moradi2021integrated,pineda2018chronological,zhang2018coordinated,dvorkin2017co,moradi2021deep}. In fact, according to the international energy agency (IEA) report, the two-degrees global warming scenario requires 450 GW installed ESS by 2050 \cite{olabi2021critical}.   

In a large-scale power system expansion planning model with different modern planning options, both the investment and operation of system should be evaluated, so the investment decisions rely on the temporal data used for modeling the operational details. Considering all time steps for each year in a long-term expansion planning problem will result in a practically unsolvable model. With this in mind, time series aggregation (TSA) methods can be utilized to capture the variability of load demand and RES output power in an expansion planning model \cite{teichgraeber2022time}. By using TSA, the temporal data are aggregated into some limited representative time periods, aiming to reduce the temporal complexity of long-term expansion models while capturing the short-term variability as much as possible. An important challenge in the representative time period extraction process and optimization modeling is incorporating inter-period cycles of long-term ESSs that can take several weeks or months, like pumped hydroelectric and power-to-gas storage technologies. This challenge is usually ignored when only short-term battery ESSs (BESSs) with intra-day cycles are considered. In this regard, a proper method is required to precisely capture the chronology of input data through the final extracted representative time periods, and to efficiently represent these in the optimization model. In addition to the chronology of data, extreme values or transitions in the input data that drive the constraints of the expansion planning model must be preserved by the TSA method \cite{gao2021spectral}.

In \cite{zhang2018coordinated} a k-means clustering algorithm, and in \cite{dvorkin2017co} a hierarchical clustering method is used to extract some representative days (RDs) for operational variability capturing in transmission expansion planning (TEP) model considering BESS. Neither the k-means nor the hierarchical clustering method directly captures the chronology between representative time periods. In this regard, in \cite{pineda2018chronological} a chronological time-period clustering (CTPC) method, which is mainly based on the agglomerative hierarchical Ward's method \cite{ward1963hierarchical}, is presented. CTPC can capture the chronology of time periods by iteratively comparing and merging the adjacent clusters. In both \cite{moradi2021integrated} and \cite{moradi2021deep} CTPC is utilized to capture the uncertainties of load demand and RES considering short-term BESS. In \cite{moradi2021integrated} an integrated TEP and generation expansion planning structure, and in \cite{moradi2021deep} a hurricane-resilient TEP model is proposed. In \cite{moradi2021deep} and \cite{dominguez2021multi} the CTPC method is improved in terms of capturing more data chronology and extreme values by developing the algorithm as a multi-stage procedure. A multi-stage CTPC (MCTPC) algorithm is proposed in \cite{dominguez2021multi} to capture different seasonality of time series in power system expansion studies. However, the proposed MCTPC in \cite{dominguez2021multi} is effective mostly for time series with a certain daily, weekly, or monthly pattern like load demand. To improve the effectiveness of CTPC in terms of capturing extreme values, a priority-based CTPC (PCTPC) method was introduced in \cite{garcia2022priority} in which extreme values are defined as maximum values of load and minimum values of RESs. However, there is no guarantee that the extracted extreme values in \cite{garcia2022priority} are the maximum values of net-load, which are expected to drive generation investments to avoid load shedding. In \cite{gao2021spectral} a demand-oriented spectral clustering method is presented for extracting RDs considering net-load and ramp of net-load duration curve as feature vectors. Another solution to capture the chronology of data is utilizing system states \cite{orgaz2022modeling}. By considering extracted system states, different hours with similar features can be compacted in the same sequential time periods based on the closest system state. This procedure leads to a high computational burden due to the high number of sequential time periods \cite{gonzato2021long}. In \cite{gonzato2021long} several TSA methods are compared and it is concluded that representative extraction method and model formulations are important in capturing the chronology of data.

Based on previous studies, the main research gaps can be mentioned as the lack of a TSA method with adaptability to the optimization model that can capture both extreme values of net-load and temporal dynamic chronology of data. In reality, extreme values are the maximum difference between load demand and RES output power that can lead to load shedding. Therefore, extreme values are very important to consider adequacy in power systems expansion planning. In addition, temporal dynamic chronology is essential to model operational flexibility under penetration of RES and ESS with intra and inter-day cycles. In contrast, in the majority of previous research works, just the chronology inside each representative time period is considered \cite{zhang2018coordinated,dvorkin2017co,pineda2018chronological,moradi2021integrated,moradi2021deep,dominguez2021multi,garcia2022priority,gao2021spectral}. With this in mind, long-term ESS with seasonal charging and discharging cycles are often not incorporated in expansion planning models. 

Therefore, in this paper a hybrid clustering-based TSA algorithm is proposed to capture both extreme values and temporal dynamics of input data for a complex co-planning model. The clustering method used to extract representative days is sensitive to extreme values, specifically the maximum value of net-load (difference between load demand and RES output power). In addition, the temporal dynamic chronology of data is captured by developing a mapping process to make sequential blocks of real time periods linked to the same closest extracted representative time periods. Then, a co-planning model of transmission lines, wind power plants (WPPs), short-term battery and long-term pumped hydroelectric energy storage systems (BESS and PHESS) is demonstrated to evaluate the effectiveness of proposed method. In the developed co-planning model, the ESS formulation is adapted to the proposed representation of linked representative days. Finally, a comparative analysis is presented by implementing some methodologies in recently published studies.

\section{Capturing Operational Variability}
\begin{table}[!b]
\renewcommand{\arraystretch}{1.2}
\caption{Clustering-Based  Time Series Aggregation Algorithm}
\label{table1}
\centering
\phantom{~}\noindent
\begin{tabular}{p{0.46\textwidth}}
\hline\hline
\textbf{Start}: Inputs: Load and RES data, and the number of representative days (NRD).\\
\hline\hline
\textbf{1.} Consider the input data at each day as initial clusters, i.e., $rd=365$, \\
\textbf{2.} Calculate the centroid of each cluster $rd$, i.e., $\bar{\mathbf{x}}_{rd}$,  as  \eqref{eq:001}.\\
\textbf{3.} Measure dissimilarity of all clusters $rd$ and $rd^{\prime}$ based on \eqref{eq:002}.\\
\textbf{4.} Merge two clusters $\bar{rd}$ and $\bar{rd^{\prime}}$ with the minimum dissimilarity based on the following steps:\\
\textbf{If} \ One of $\bar{rd}$ and $\bar{rd^{\prime}}$ contains an extreme day, \\
\textbf{4.1.} Select the extreme day (or pick one randomly if both clusters contain extreme days) and copy its features to all days in the combined cluster, to ensure its centroid is equal to the extreme day. \\
 
\textbf{Else} \\ 
\textbf{4.2.} \  Merge the two clusters, by updating cluster labels for each day. \ \textbf {End If} \\
\textbf{5.} $rd \leftarrow rd - 1$ \\ 
\textbf{If} $rd=NRD$ \\ 
\textbf{5.1} Go to step 6 \\ \textbf{Else} \\
\textbf{5.2} Return to step 2 \\ \textbf {End If} \\
\textbf{6.} Calculate the final representative days using the related centroids $\bar{\mathbf{x}}_{rd}$, and consider the number of days in each cluster as final weights $\rho_{rd}$.\\
\textbf{7.} Compare the centroids of real days with all $\bar{\mathbf{x}}_{rd}$ and map each real day with the closest $\bar{\mathbf{x}}_{rd}$.\\
\textbf{8.} Create SLDs by merging adjacent days that are mapped to the same RDs.\\
\textbf{9.} Create mapping matrix $MP_{sld}^{rd}$, and each SLD's weight $\rho_{sld}$.\\
\hline\hline
\textbf{End}: Outputs: $\bar{\mathbf{x}}_{rd}$, $\rho_{rd}$, $MP_{sld}^{rd}$, and $\rho_{sld}$\\
\hline\hline
\end{tabular}
\end{table}
The \emph{intra}day dynamics of short-term ESS can be modeled using RDs. To also consider the \emph{inter}-day dynamics of ESSs, especially for long-term storage, the chronological sequence of RDs and their impact on stored energy should be modeled. This section describes the procedure to extract sequences of representative days. First, a hierarchical clustering method is used to identify representative days, based on the centroid linkage criterion \cite{teichgraeber2022time}. 
Clusters are represented by their centroids, given by
\begin{equation} \label{eq:001}
    \boldsymbol{\overline{\mathbf{x}}_A=\frac{1}{|A|} \sum_{i \in A} \mathbf{x}_i},
\end{equation} 
where \emph{A} is the label of the cluster that consists of $|A|$ elements $x_i$, labeled by $i$. The dissimilarity metric between two clusters \emph{A} and \emph{B}, with centroids $\overline{\mathbf{x}}_A$ 
and $\overline{\mathbf{x}}_B$
is calculated based on Ward's method \cite{pineda2018chronological}. 

\begin{equation} \label{eq:002}
D(A, B)=\frac{2|A||B|}{|A|+|B|}\left\|\overline{\mathbf{x}}_A-\overline{\mathbf{x}}_B\right\|^2
\end{equation}
In addition to load demand and RES output power data, net-load data are incorporated as feature vectors to place more emphasis on capturing net load. In addition, to avoid flattening of extremes, the day with the highest net load is marked as an \emph{extreme} day, and will act as the centroid for its cluster, overriding \eqref{eq:001}. 
The overall structure of proposed clustering algorithm is described in Table \ref{table1}. 

Representative days capture the chronology within each day, but not between days. 

Therefore, to capture the temporal dynamics of the data, blocks of sequentially linked days (SLDs) are identified as real days linked to the same closest RD. 

\begin{figure}[!t]
\centering
\includegraphics[scale=0.28]{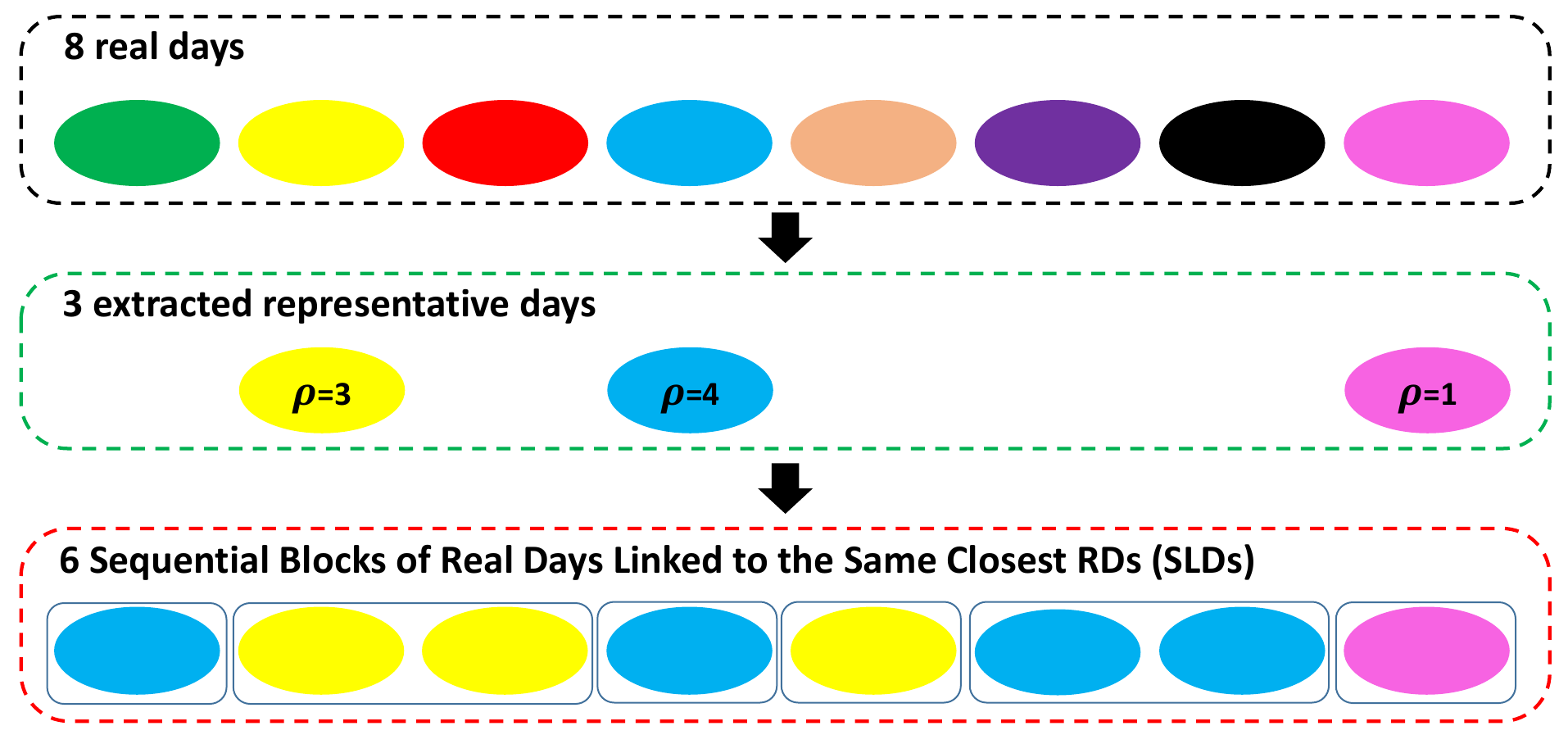}\caption{Concept of mapping process to construct sequential blocks of real days linked to the same closest representative days}
\label{fig:d01}
\end{figure}

The mapping process to construct SLDs is illustrated in Fig.~\ref{fig:d01}. In this example, three RDs are extracted from eight real days. The weight ($\rho$) of each RD is shown. Then, six SLDs are constructed using a mapping process to preserve the chronology of data as much as possible. Note that there is a relation between the number of RDs and SLDs. In the long-term storage model, only state-of-charge variables will be associated with each SLD. 

Accordingly, with an acceptable computational time, the full space model can be precisely approximated by several RDs and SLDs. In Table \ref{table1}, the mapping process is implemented in steps 7 to 9.
\section{Proposed Co-Planning Formulation}
In this section, the proposed co-planning model is formulated as a MILP model according to DC optimal power flow (DC-OPF). The problem is modeled as a static or single-year co-planning to evaluate the proposed clustering-based TSA algorithm. The objective function and all techno-economical constraints are presented in \eqref{eq:003}, \eqref{eq:004}, and \eqref{eq:005}-\eqref{eq:039}.
\begin{subequations}
\begin{equation} \label{eq:003}
\begin{aligned}
&\operatorname{Min} Z={TC+TO}&&\\ \\
&{TC}=\sum_{l \in \Omega_{nl}} \left[ (IC_l+Rw_l) \cdot LL_l \cdot Y_{l}\right] +\sum_{i \in \Omega_{s b}} \left[Cs_i\cdot S_{i} +Cc_i \cdot C_{i}\right]&&\\
&+\sum_{i \in \Omega_W} \left[ICw_i \cdot Pw_{i}\right]&& 
\end{aligned}   
\end{equation} 
\begin{equation}\label{eq:004} 
\begin{aligned}
&{TO}=\frac{1}{(1+r)} \times\left[\sum_{\boldsymbol{rd} \in \Omega_{\boldsymbol{RD}}} \rho_{rd} \times \sum_{\boldsymbol{h} \in \Omega_{\boldsymbol{H}}}\left[\sum_{i \in \Omega_G} \left[\right.\left(\chi_i \cdot R_{i,rd,h}\right) \right.\right. &&\\
&+\sum_{\mathcal{P} \in \Omega_\mathbb{P}} \left(Cg_i^\mathcal{P} \cdot Ps_{i,rd,h,\mathcal{P}}\right)\left.\right] \left.\left. +\sum_{i \in \Omega_w} Cwc_i \cdot PC_{i,rd,h}\right]\right]&&
\end{aligned}   
\end{equation} 
\end{subequations}
\begin{equation} \label{eq:005}
\begin{aligned}
&P_{i,rd,h}+\left[Wf_{rd}^{h} \cdot P w_{i}-PC_{i,rd,h}\right]+\left[Pd_{i,rd,h}-Pc_{i,rd,h}\right]&\\&
-\sum_{l \in \Omega_{el}} A_i^l \cdot Pe_{l,rd,h}-\sum_{l \in \Omega_{nl}} K_i^l \cdot Pl_{l,rd,h}=\left((1+Lg) \cdot Lf_{rd}^{h} \cdot Ld_i^{PK}\right)&\\&
\qquad \qquad \qquad \qquad \qquad \qquad \qquad \forall i \in \Omega_B, rd \in \Omega_{RD}, h \in \Omega_H
\end{aligned}
\end{equation}
\begin{equation} \label{eq:006}
\begin{aligned}
&0\leq P_{i,rd,h} \leq P_i^{\max } \cdot I_{i,rd,h} \quad \forall i \in \Omega_G, rd \in \Omega_{RD},h \in \Omega_{H}&& 
\end{aligned}
\end{equation}
\begin{equation} \label{eq:007}
\begin{aligned}
&P_{i,rd,h}=\sum_{\mathcal{P}=1}^\mathbb{P} Ps_{i,rd,h,\mathcal{P}} \quad \forall i \in \Omega_G, rd \in \Omega_{RD},h \in \Omega_{H}&& 
\end{aligned}
\end{equation}
\begin{equation} \label{eq:008}
\begin{split}
0 \leq Ps_{i,rd,h,\mathcal{P}} \leq P_i^{\max } \cdot I_{i,rd,h}/\mathbb{P} \ \ \forall i \in \Omega_G, rd \in \Omega_{RD}, h \in \Omega_{H},\\ \mathcal{P} \in \Omega_\mathbb{P}
\end{split}
\end{equation}
\begin{equation}\label{eq:009}
\begin{aligned}
&0 \leq Pw_{i} \leq Pw_i^{\max } \quad \forall i \in \Omega_w&&
\end{aligned}
\end{equation}
\begin{equation}\label{eq:010}
\begin{aligned}
&\alpha\% \times(1+L g) \times \sum_{i \in \Omega_B} Ld_i^{pk} \leq \sum_{i \in \Omega_w} Pw_{i}&&
\end{aligned}
\end{equation}
\begin{equation}\label{eq:011}
\begin{aligned}
&0 \leq PC_{i,rd,h} \leq Wf_{rd}^{h} \cdot Pw_{i} \quad \forall i \in \Omega_w, rd \in \Omega_{RD}, h \in \Omega_H&&
\end{aligned}
\end{equation}
\begin{equation}\label{eq:012}
\begin{aligned}
&\sum_{i \in \Omega_w} \sum_{rd \in \Omega_{RD}} \sum_{h \in \Omega_H} PC_{i,rd,h} \leq \beta\% \times \sum_{i \in \Omega_w} \sum_{rd \in \Omega_{RD}} \sum_{h \in \Omega_H} Wf_{rd}^{h} \cdot Pw_{i}&&
\end{aligned}
\end{equation}
\begin{equation}\label{eq:015}
\begin{aligned}
&0 \leq R_{i,rd,h} \leq P_{i,rd,h} \quad \forall i \in \Omega_G, rd \in \Omega_{RD}, h \in \Omega_H&&
\end{aligned}
\end{equation}
\begin{equation}\label{eq:016}
\begin{aligned}
&R_{i,rd,h}+P_{i,rd,h} \leq P_i^{\max} \quad \forall i \in \Omega_G, rd \in \Omega_{RD}, h \in \Omega_H&&
\end{aligned}
\end{equation}
\begin{equation}\label{eq:017}
\begin{aligned}
&\sum_{i \in \Omega_G} R_{i,rd,h} \geq(\sigma\%) \times \sum_{i \in \Omega_w} Wf_{rd}^{h} \cdot Pw_{i}&&\\
&+(\mu\%) \times (1+Lg) \cdot Lf_{rd}^{h} \cdot \sum_{i \in \Omega_B} Ld_i^{p k} \quad \forall rd \in \Omega_{RD}, h \in \Omega_H&&
\end{aligned}
\end{equation}
\begin{equation}\label{eq:018}
\begin{aligned}
&Pe_{l,rd,h}-\sum_{i \in \Omega_B} \Psi \cdot B_l \cdot A_i^l \cdot \theta_{i,rd,h}=0 \ \forall l \in \Omega_{e l}, rd \in \Omega_{RD}, h \in \Omega_H &&
\end{aligned}
\end{equation}
\begin{equation}\label{eq:019}
\begin{aligned}
&-P_l^{\max}\leq Pe_{l,rd,h} \leq P_l^{\max} \quad\forall l \in \Omega_{el}, rd \in \Omega_{RD}, h \in \Omega_H &&
\end{aligned}
\end{equation}
\begin{equation}\label{eq:020}
\begin{aligned}
-M_l \cdot\left(1-Y_{l}\right) \leq Pl_{l,rd,h}-\sum_{i \in \Omega_B} \Psi \cdot B_l \cdot K_i^l \cdot \theta_{i,rd,h} \leq M_l \cdot\left(1-Y_{l}\right)\\
\forall l \in \Omega_{nl}, rd \in \Omega_{RD}, h \in \Omega_H
\end{aligned}
\end{equation}
\begin{equation}\label{eq:021}
\begin{aligned}
-P_l^{\max} \cdot Y_{l} \leq Pl_{l,rd,h} \leq P_l^{\max} \cdot Y_{l} \ \ \forall l \in \Omega_{nl}, rd \in \Omega_{RD}, h \in \Omega_H
\end{aligned}
\end{equation}
The objective function is the minimization of two terms. The first term is total capital cost (\emph{TC}) which consists of new transmission line, ESS, and WPP investment cost defined in \eqref{eq:003}. \emph{IC, Rw, LL} and \emph{Y} are new line \emph{l} investment and right-of-way costs (\$/km), length (km), and binary construction variable, respectively. \emph{Cs, Cc, S} and \emph{C} are ESS energy and power capacity investment cost (\$/MWh and \$/MW) and variable in each bus \emph{i}, respectively. \emph{ICw} and \emph{Pw} are investment cost (\$/MW) and total power capacity of new WPP in each bus \emph{i}. The second term is total operation cost (\emph{TO}) in which hourly thermal generation units flexible spinning reserve and operation cost, along with wind curtailment cost are minimized as defined in \eqref{eq:004}. Note that \emph{r} is the effective interest rate used to calculate the discounted present value of \emph{TO} accrued throughout the year. \emph{rd} and \emph{h} are representative days and the hours inside days, and $\mathcal{P}$ is the number of segments used for linearizing thermal units generation cost function. $\rho$, $\chi$, \emph{Cg} and \emph{Cwc} are weight of each RD (or SLD), flexible spinning reserve cost, power generation cost in each segment, and wind curtailment cost, respectively. Flexible spinning reserve, power generation of each segment, and amount of curtailed wind variables are presented with \emph{R, Ps} and \emph{PC}, respectively. The power balance for all system buses is defined in \eqref{eq:005} in which \emph{P, Pd, Pc, Pe} and \emph{Pl} are variables of power output of generation units, charging and discharging power of ESS, transferred power through existing and new lines, respectively. \emph{Wf, Lf, Lg, Ld, A} and \emph{K} are wind and load representative factors, load growth, peak load, directional connectivity matrices of existing and new lines, respectively. The thermal units constraints considering on/off status variable \emph{I} are defined in \eqref{eq:006}-\eqref{eq:008}. To model WPP penetration considering the probable wind energy curtailment, constraints in \eqref{eq:009}-\eqref{eq:012} are defined. The expected WPP share in supplying the load and the maximum wind curtailment are considered by $\alpha$ and $\beta$. In \eqref{eq:015}-\eqref{eq:017} flexible spinning reserve modeling is formulated. The minimum bound of total
hourly flexible spinning reserve is considered as 5\% of the total expected WPP power plus 3\% of the system load \cite{moradi2021deep}. In \eqref{eq:018}-\eqref{eq:019} existing and in \eqref{eq:020}-\eqref{eq:021} new transmission lines constraints are presented. $\Psi$, \emph{B, M} and $\theta$ are the base power, per unit susceptance of lines, big-M, and voltage angle variable, respectively.

To model the charging and discharging cycles along with stored energy in ESS, the formulations for ESS are designed in such a way as to be adapted to the proposed clustering algorithm considering RDs and SLDs. The (dis-)charging behaviour of each ESS is optimized for each given RD, and repeated across days within an SLD. Energy balance is tracked across SLDs within the year, based on aggregate stored energy within each RD. The formulations for ESS modeling are presented in \eqref{eq:022}-\eqref{eq:039}. Note that some formulations, e.g., \eqref{eq:028}, \eqref{eq:0282}, \eqref{eq:038} and \eqref{eq:039}, are derived from the formulations presented in \cite{gonzato2021long} to model storage using enhanced representative days (ERD). The main difference is that, by using SLDs, the number of variables for state of charge is reduced, addressing the main drawback of the method, at the expense of disallowing `mixtures' of RDs.   
\begin{equation}\label{eq:022}
\begin{aligned}
&&0 \leq \eta c \cdot Pc_{i,rd,h} \leq C_i \quad \forall i \in \Omega_{sb},rd \in \Omega_{RD},h \in \Omega_{H}&&
\end{aligned}
\end{equation}
\begin{equation}\label{eq:023}
\begin{aligned}
&0 \leq 1 / \eta d \cdot Pd_{i,rd,h} \leq C_i \quad \forall i \in \Omega_{sb},rd \in \Omega_{RD},h \in \Omega_{H}&
\end{aligned}
\end{equation}
\begin{equation}\label{eq:024}
\begin{aligned}
&0 \leq C_i \leq C_{i}^{\max}, \ \ 0 \leq S_i \leq S_{i}^{\max}, \ \ C_i \cdot \varphi \leq S_i \quad \forall i \in \Omega_{sb}&&
\end{aligned}
\end{equation}
\begin{equation}\label{eq:025}
\begin{aligned}
&\eta c . Pc_{i,rd,h} \leq U_{i,rd,h} \cdot C_{i}^{\max} \quad \forall i \in \Omega_{sb},rd \in \Omega_{RD},h \in \Omega_{H}&
\end{aligned}
\end{equation}
\begin{equation}\label{eq:026}
\begin{split}
1/\eta d \cdot Pd_{i,rd,h} \leq\left(1-U_{i,rd,h}\right) \cdot C_{i}^{\max} \quad \forall i \in \Omega_{sb},rd \in \Omega_{RD},\\ h \in \Omega_{H}
\end{split}
\end{equation}
\begin{equation}\label{eq:027}
\begin{split}
\lambda_{i,rd,h}=\lambda_{i,rd,h-1}+\left(\eta c \cdot Pc_{i,rd,h}-1 /\eta d \cdot Pd_{i,rd,h}\right) \quad \forall i \in \Omega_{sb},\\rd \in \Omega_{RD},h \in \Omega_{H_{2:end}}
\end{split}
\end{equation}
\begin{equation}\label{eq:028}
\begin{aligned}
\lambda_{i,rd,h_{1}}=\lambda Z_{i,rd}+\left(\eta c \cdot Pc_{i,rd,h_{1}}-1 /\eta d \cdot Pd_{i,rd,h_{1}}\right) \quad \forall i \in \Omega_{sb},\\rd \in \Omega_{RD}
\end{aligned}
\end{equation}
\begin{equation}\label{eq:0281}
\begin{aligned}
&0 \leq \lambda_{i,rd,h}\leq S_i \quad \forall i \in \Omega_{sb},rd \in \Omega_{RD},h \in \Omega_{H}&
\end{aligned}
\end{equation}
\begin{equation}\label{eq:0282}
\begin{aligned}
&0 \leq \lambda Z_{i,rd}\leq S_i \quad \forall i \in \Omega_{sb},rd \in \Omega_{RD}&
\end{aligned}
\end{equation}
\begin{equation}\label{eq:029}
\begin{aligned}
\Delta \lambda_{i,rd}=\sum_{h \in \Omega_{H}} \left(\eta c \cdot Pc_{i,rd,h}-1/\eta d \cdot Pd_{i,rd,h}\right) \ \forall i \in \Omega_{sb},\\ rd \in \Omega_{RD}
\end{aligned}
\end{equation}
\begin{equation}\label{eq:030}
\begin{split}
E_{i,sld}=E_{i,sld-1}+\rho_{sld}\times\sum_{rd\in\Omega_{RD}} MP_{sld}^{rd} \times \Delta \lambda_{i,rd}\quad \forall i \in \Omega_{sb},\\ sld \in \Omega_{SLD_{2:end}} 
\end{split}
\end{equation}
\begin{equation}\label{eq:031}
\begin{aligned}
E_{i,sld_{1}}\leq E_{i,sld_{end}}+\rho_{sld_{1}}\times\sum_{rd \in \Omega_{RD}} MP_{sld_{1}}^{rd} \times \Delta \lambda_{i,rd} \quad \forall i \in \Omega_{sb} 
\end{aligned}
\end{equation}
\begin{equation}\label{eq:034}
E_{i,sld}+\sum_{rd \in \Omega_{RD}} MP_{sld}^{rd} \times (\lambda_{i,rd}^{\max }-\Delta \lambda_{i,rd}) \leq S_i \ \ \forall i \in \Omega_{sb},\\ sld \in \Omega_{SLD}
\end{equation}
\begin{equation}\label{eq:035}
\begin{aligned}
&E_{i,sld-1}-\sum_{rd \in \Omega_{RD}} MP_{sld}^{rd} \times \lambda_{i,rd}^{\min } \geq 0  \ \ \quad \forall i \in \Omega_{sb},sld \in \Omega_{SLD} &&
\end{aligned}
\end{equation}
\begin{equation}\label{eq:036}
\begin{aligned}
&E_{i,sld-1}+\sum_{rd \in \Omega_{RD}} MP_{sld}^{rd} \times \lambda_{i,rd}^{\max } \leq S_i \ \  \quad \forall i \in \Omega_{sb},sld \in \Omega_{SLD} &&
\end{aligned}
\end{equation}
\begin{equation}\label{eq:037}
E_{i,sld}-\sum_{rd \in \Omega_{RD}} MP_{sld}^{rd} \times (\lambda_{i,rd}^{\min }-\Delta \lambda_{i,rd}) \geq 0 \ \forall i \in \Omega_{sb},\\ sld \in \Omega_{SLD}
\end{equation}
\begin{equation}\label{eq:038}
\begin{aligned}
&0 \leq \lambda_{i,rd}^{\min} \geq \lambda Z_{i,rd}-\lambda_{i,rd,h} \quad \forall i \in \Omega_{sb},rd \in \Omega_{RD},h \in \Omega_{H}&
\end{aligned}
\end{equation}
\begin{equation}\label{eq:039}
\begin{aligned}
&0 \leq \lambda_{i,rd}^{\max}\geq \lambda_{i,rd,h}-\lambda Z_{i,rd}\ \ \forall i \in \Omega_{sb},rd \in \Omega_{RD},h \in \Omega_{H}&
\end{aligned}
\end{equation}
\begin{figure}[!b]
\centering
\includegraphics[scale=0.388]{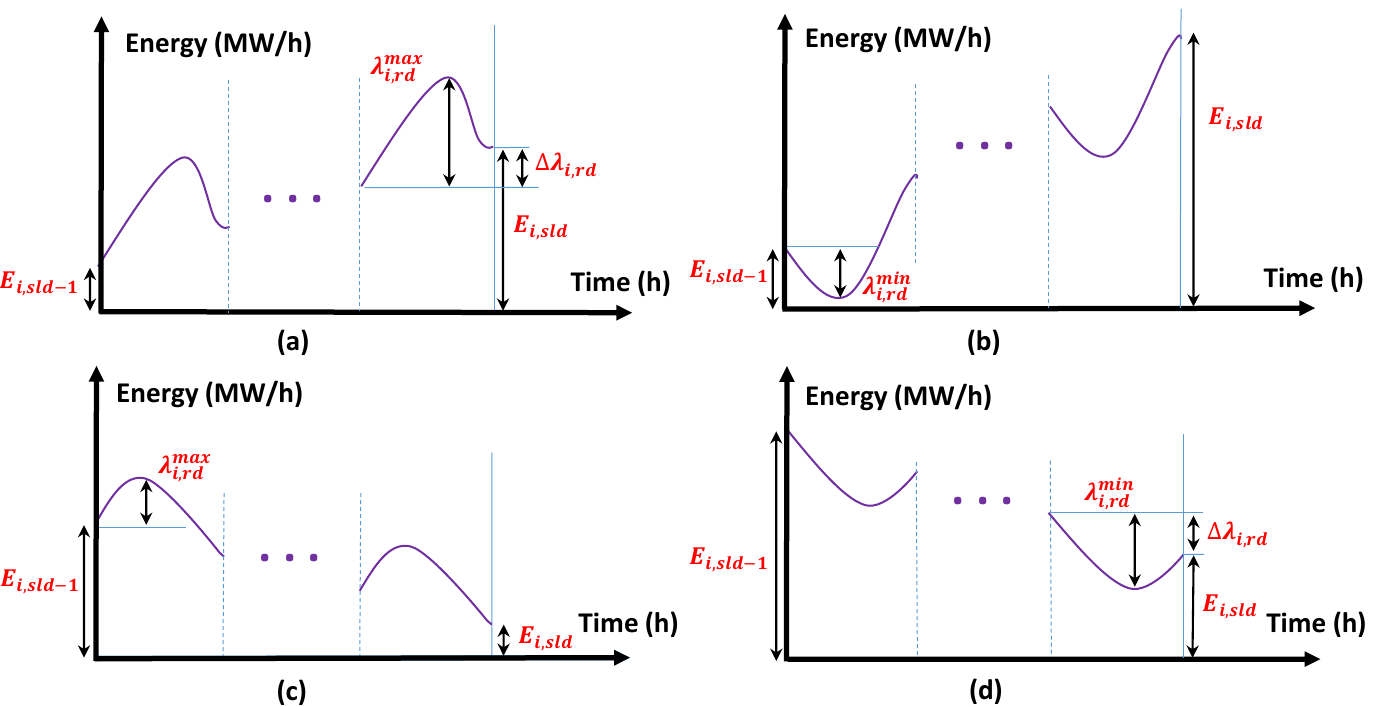}
\caption{An example of ESS stored energy levels inside each SLD, with bounds \eqref{eq:034}-\eqref{eq:037} illustrated in (a)-(d), respectively}
\label{fig:d02}
\end{figure}
In \eqref{eq:022} and \eqref{eq:023} the charging and discharging power of each  ESS are bounded with regard to the installed power capacity and efficiency ($\eta$). The limits of power and energy capacity along with the energy-to-power ratio ($\varphi$) of ESS are defined in \eqref{eq:024}. In constraints \eqref{eq:025} and \eqref{eq:026} the charging or discharging status of ESS in each hour is determined by the binary variable of \emph{U}. The hourly stored energy levels inside each day considering the initial values are calculated in \eqref{eq:027} and \eqref{eq:028}. $\lambda$ and $\lambda Z$ are positive variables bounded in \eqref{eq:0281} and \eqref{eq:0282}, respectively. The final stored energy level for each day is defined in \eqref{eq:029} and the stored energy level for each SLD (i.e., \emph{E}) is calculated in \eqref{eq:030} in which \emph{MP} is the mapping matrix. The state of charge for the first SLD is related to the final SLD by \eqref{eq:031}. Equations \eqref{eq:034}-\eqref{eq:037} ensure that the stored energy never exceeds the lower and upper bounds during each SLD.  This makes use of the minimum and maximum (relative) energy levels, defined as positive variables $\lambda^{min}$ and $\lambda^{max}$ in \eqref{eq:038} and \eqref{eq:039}, respectively. These bounds are graphically depicted in Fig.~\ref{fig:d02}.

\section{Numerical Results}
\begin{figure}[!b]
\centering
\includegraphics[scale=0.410]{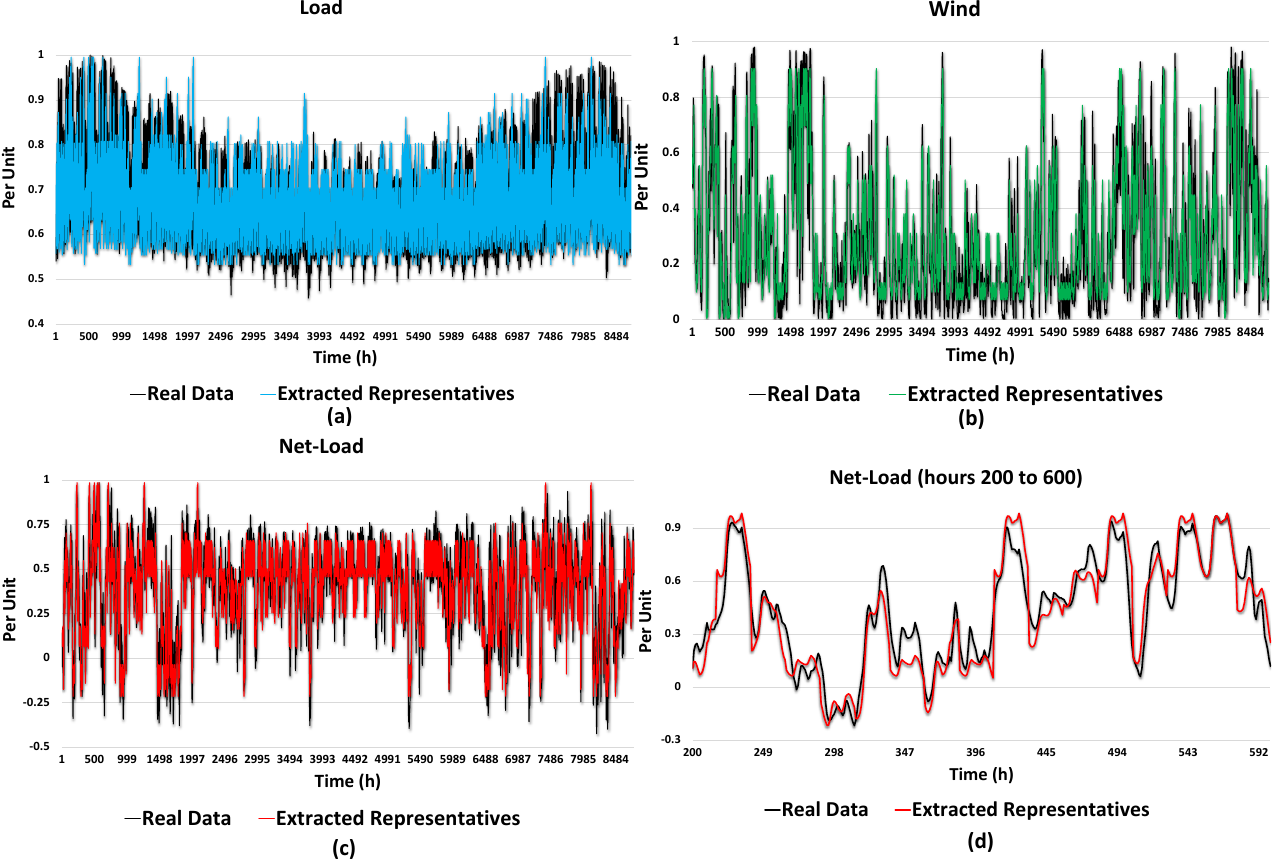}
\caption{Extracted representatives utilizing the proposed method for: (a) load, (b) wind, (c) net-load, and (d) net-load of hours 200 to 600}
\label{fig:d03}
\end{figure}

In this section the simulation results are presented. First, the developed clustering-based method is implemented on the historic load demand \cite{web1}, and wind power \cite{renewables.ninja} of the Netherlands in 2019. Fourteen RDs, each with 24 hours, are extracted with the proposed method that have led to 288 SLDs as illustrated in Fig. \ref{fig:d03}. In this figure, the representatives are aggregated across the year considering $\rho_{sld}$.
Second, a modified 6-bus Garver’s test system is used to test the effectiveness of proposed algorithm for extracting proper representatives for a co-planning model. This system has five existing buses and a new bus that can be connected to the system through buses 2 and 4. A new bus 7 is also considered as an expansion bus including a WPP. In buses 3, 5, and 7, WPPs with the capacity of 150, 150, and 300 MW can be installed, respectively. Furthermore, six lines exist in the system and eleven candidate lines are considered as expansion options. Two PHESS in buses 3 and 7, along with three BESS in buses 1, 4, and 5 are assumed as ESS candidates. The BESS technology characterized by $\eta_c$ and $\eta_d$ both equal to 0.9, $\varphi$ equal to 4 hours, and the cost of 500 \$/KW and 50 \$/KWh. For considered PHESS, $\eta_c$ and $\eta_d$ both equal to 0.548, $\varphi$ equals to 1000 hours, and the costs are 30 \$/KW and 1.5 \$/KWh. All characteristics and costs of ESS are extracted from \cite{gonzato2021long}.

\begin{figure}[!b]
\centering
\includegraphics[scale=0.365]{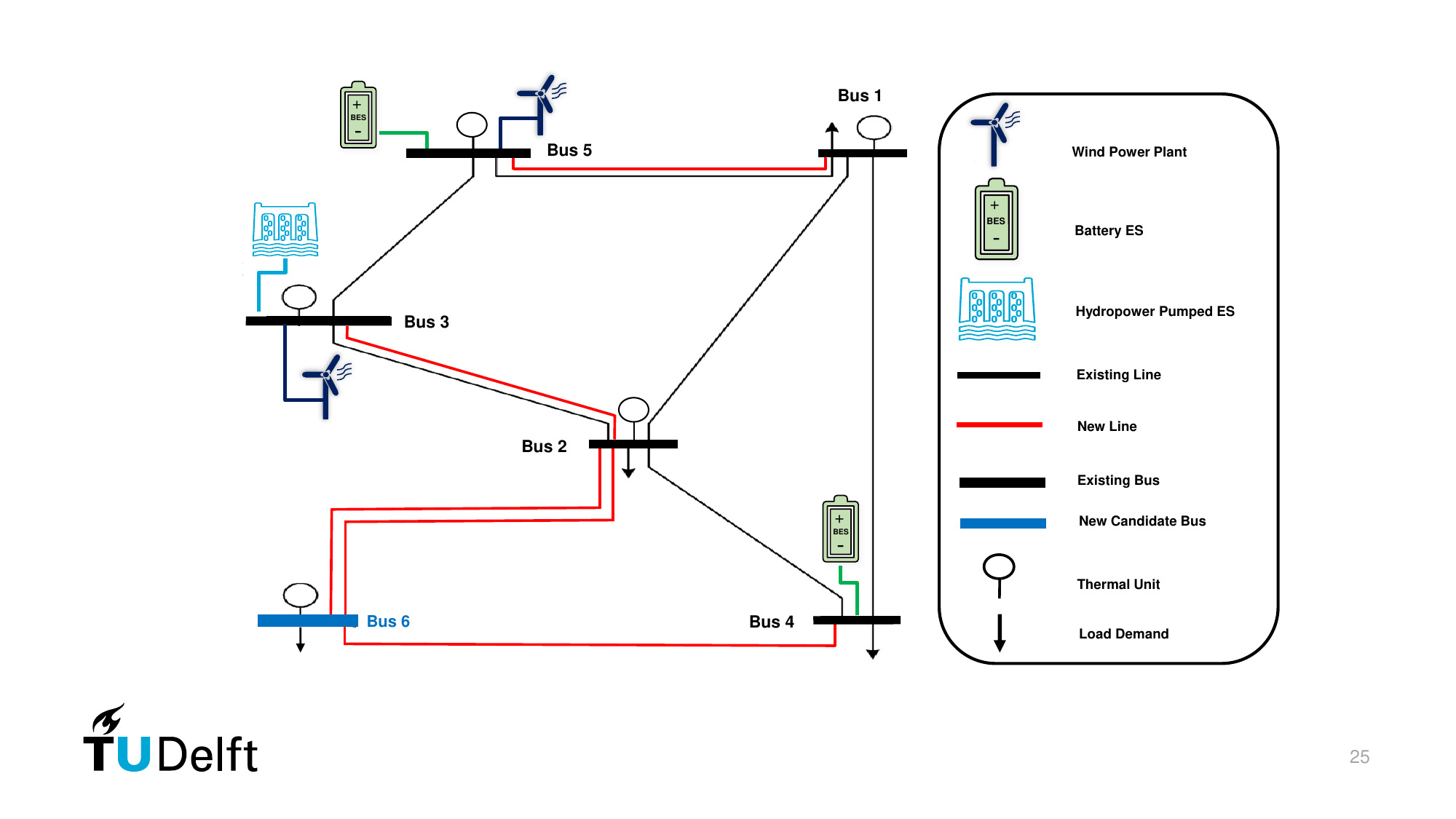}
\caption{Illustration of obtained results for the benchmark case}
\label{fig:d04}
\end{figure}

The \emph{IC} and \emph{Rw} for transmission lines are 1, and 0.04 M\$/Km, and yearly load growth along with interest rate are both 5\% \cite{moradi2020hybrid}. The \emph{ICw} is assumed as 2 M\$/MW \cite{moradi2021deep}. Load and generation is assumed to be perfectly correlated in all buses. \begin{table}[!t]{}
\renewcommand{\arraystretch}{1.11}
\caption{The Obtained Results for the Benchmark Case}
\label{table2}
\centering
\phantom{~}\noindent
\begin{tabular}{c|c||c}
\hline \hline \multicolumn{2}{c||}{$\mathbf{Planning \ Options:}$} & $\mathbf{Location:}$ \\
\hline \hline \multicolumn{2}{c||}{ $\mathbf{Line}$ } & $1-5,2-3,2 \times(2-6), \& \ 4-6$ \\
\hline \multirow{2}{*}{$\mathbf{ESS}$} & $\mathbf{BESS}$ & Buses 4 \& 5 \\
\cline { 2 - 3 } & $\mathbf{PHESS}$ & Bus 3 \\
\hline \multicolumn{2}{c||}{ $\mathbf{WPP}$ } & Buses 3 \& 5 \\
\hline \hline \multicolumn{3}{c}{$\mathbf{Costs}$ $(\times \mathbf{10}^6 $\$$):$}  \\
\hline \hline \multicolumn{2}{c||}{$\mathbf{TC:}$} & $155.51$ \\
\hline \multicolumn{2}{c||}{$\mathbf{TO:}$} & $63.23$ \\
\hline \multicolumn{2}{c||}{$\mathbf{TESSC:}$} & $15.01$ \\
\hline \multicolumn{2}{c||}{$\mathbf{Z:}$} &  $218.74$ \\
\hline \hline \multicolumn{2}{c}{$\mathbf{Wind \ Curtailment:}$} & $ 0 \ MW$ \\
\hline \multicolumn{2}{c}{$\mathbf{CPU \ Time:}$} & $ 48,660 \ s$ \\
\hline \hline
\end{tabular}
\end{table}
All input data and parameters are also accessible in \cite{Modified}. The MILP model of proposed co-planning problem was solved by CPLEX solver in the GAMS environment \cite{General} and the proposed clustering-based algorithm was coded in Matlab software \cite{matlabb} in a PC with Intel Xeon W-2223 CPU 3.60 GHz, and 16 GB of RAM. As a benchmark, the model is simulated using all hours of the year. The obtained results are presented in Table \ref{table2} and illustrated in Fig. \ref{fig:d04} for more clarification. As reported in Table \ref{table2}, \emph{Z} is 218.74 M\$ consisting of 155.51 and 63.23 M\$ of \emph{TC} and \emph{TO} costs, respectively.
As shown in Fig. \ref{fig:d04}, five new lines, two BESS and one PHESS, along with two WPPs are installed in the system. Due to ESS, no wind energy is curtailed. The CPU time for the benchmark exceeds 13 hours for this single-year co-planning model and a simple system, demonstrating the necessity of reducing the complexity of expansion models by extracting representative time periods. To evaluate the effectiveness of proposed method in this paper, three different methodologies including CTPC \cite{pineda2018chronological}, PCTPC \cite{garcia2022priority}, and MCTPC \cite{dominguez2021multi} are also implemented and the results for the same number of representative days (or hours, as appropriate) are compared with the proposed method. To analyze the obtained results of each methodology, we calculate
\begin{equation}\label{eq:040}
Error=\left|\frac{\boldsymbol{f}_c^*(\widehat{v})-\boldsymbol{f}_c^*(\boldsymbol{v}^*)}{f_c^*(v^*)}\right|
\end{equation}
where $\widehat{v}$ represents the fixed decision variables obtained from the reduced space model and $v^*$ are the decision variables of the benchmark model. The function $\boldsymbol{f}_c^*()$ represents the total cost of a given set of decisions, evaluated using all hours. It is a measure of the relative cost of implementing sub-optimal decisions in the reference model. 

\begin{figure}[!t]
\centering
\includegraphics[scale=0.497]{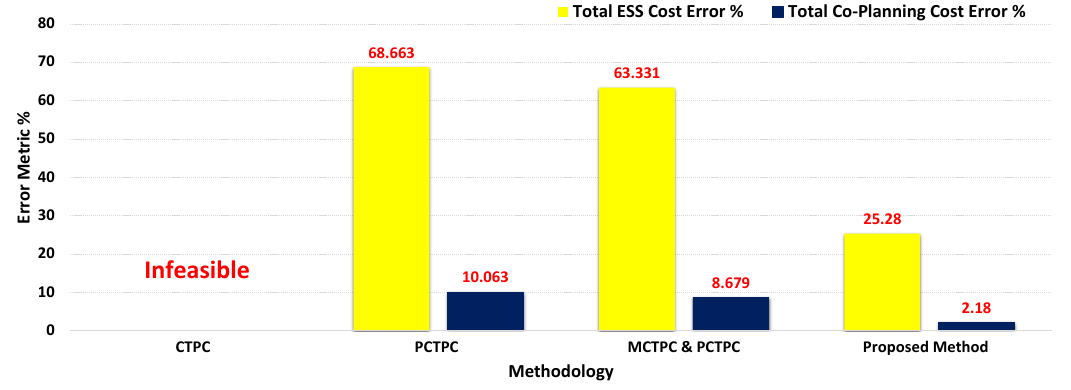}
\caption{Error comparison between different evaluated clustering-based methods for the developed co-planning model}
\label{fig:d05}
\end{figure}

As mentioned and illustrated in Fig. \ref{fig:d03}, by using the proposed method described in Table \ref{table1}, 14 RDs, each with 24 hours, are extracted and with implementing the mapping process 288 SLDs are obtained. On the other hand, 336 representative hours ($14 \times 24$) are extracted using CTPC and PCTPC methods. For MCTPC, 7 months were extracted in the first stage, and 2 days from each month in the second stage to to reach 14 RDs. The errors obtained using the proposed method, CTPC, PCTPC, and combined MCTPC and PCTPC are presented in Fig. \ref{fig:d05}, both for the total costs and for the total ESS cost. The results obtained using CTPC were operationally infeasible when all hours were used (i.e. there was a loss of load), represented as 100\%.  
This outcome confirms the importance of considering extreme values in the representative extraction process, especially when dealing with hard constraints.  
The proposed method results in significantly reduced errors, compared to both PCTPC methods. Moreover, the CPU time for the reduced space model considering the RDs extracted by the proposed model is just 279 seconds which shows a 99\% computational time saving compared to the benchmark case presented in Table \ref{table2}. In other words, by just considering 14 days, i.e., 3.8\% of all days of the year, a 99\% computational time saving can be achieved with just 2\% error in total co-planning cost. The obtained results illustrate the potential efficacy of the proposed model in providing a trade-off between computational time and accuracy considering short and long-term ESSs.
\section{Conclusion}
This paper proposed a clustering-based time series aggregation method for capturing operational variability of load and renewable output power in a developed co-planning expansion problem for transmission lines and energy storage systems considering the sitting and sizing of wind power plants. Both short-term battery and long-term pumped hydroelectric energy storage systems with intraday and interday charging and discharging cycles were incorporated into the proposed model. The main purpose was to capture both extreme values and temporal chronology of data in the final extracted representatives. The obtained result confirmed that the proposed clustering-based method is able to reduce the computational time of the developed co-planning model up to 99\% with an error of 2\% in total co-planning cost. This time saving is obtained considering just fourteen representative days which is 3.8\% of the data. By implementing a trade-off between complexity and accuracy, the error can be reduced more noticeably with more extracted representative days and consequently more sequential blocks of real days linked to the same closest representative day. Moreover, the proposed method in this paper can be utilized for more complex power systems with different types of RES.  

Future work will quantify the result quality and computational gains in a wider range of scenarios, and will investigate whether  additional efficiency can be obtained by optimizing the model representation of the representative days.

\bibliography{Reff(1)} 
\bibliographystyle{IEEEtran}
\end{document}